\documentclass{jpsj-suppl}

\usepackage{txfonts} 

\newcommand{\Slash}[1]{\ooalign{\hfil/\hfil\crcr$#1$}}

\title{Baryon resonances in the strangeness production}

\author{Ju-Jun \textsc{Xie}$^{1,2,3}$, En \textsc{Wang}$^{4}$ and Jia-Jun \textsc{Wu}$^{5}$}

\inst{$^{1}$Institute of Modern Physics, Chinese Academy of
Sciences, Lanzhou 730000, China \\
$^{2}$ State Key Laboratory of Theoretical Physics, Institute of
Theoretical Physics, Chinese Academy of Sciences, Beijing 100190,
China \\
$^{3}$ Research Center for Hadron and CSR Physics, Institute of
Modern Physics of CAS and Lanzhou University, Lanzhou 730000, China
\\
$^{4}$ Department of Physics, Zhengzhou University, Zhengzhou, Henan
450001, China \\
$^5$ Special Research Center for the Subatomic Structure of Matter
(CSSM), School of Chemistry and Physics, University of Adelaide
Adelaide 5005, Australia

}

\email{xiejujun@impcas.ac.cn}

\recdate{\today}

\abst{We have studied the $N^*(2120)$, $\Delta^*(1940)$, and the
possible $\Sigma^*(1380)$ resonances in the $\gamma p \to K^+
\Lambda(1520)$, $pp \to n K^+ \Sigma(1385)$, and $\Lambda p \to
\Lambda p \pi^0$ reactions within the resonance model and the
effective Lagrangian approach. It is shown that when the
contributions from these baryonic states were considered, the
current experimental measurement could be well reproduced. In
addition, we also demonstrate that the angular distributions provide
direct information of these reactions, which could be useful for the
investigation of those states and may be tested by future
experiments.}

\kword{Baryon resonances, effective Lagrangian approach, strangeness
production}

\begin{document}
\maketitle

\section{Introduction}

The associate production of hadrons by photon and hadron beams has
been extensively studied since it provides an excellent tool to
learn the details of the hadron spectrum. In addition, those
reactions require the creation of an $\bar{s}s$ quark pair. Thus, a
thorough and dedicated study of the strangeness production mechanism
in those reactions has the potential to gain a deeper understanding
of the interaction among strange hadrons and also the nature of the
baryon resonances. In particular, the $\gamma p \to K^+
\Lambda(1520)$ reaction is an efficient isospin $1/2$ filter for
studying nucleon resonances decaying to $K\Lambda(1520)$. As a
consequence, the experimental database on this reaction has expanded
significantly in recent years. For the $pp \to nK^+\Sigma^+(1385)$
reaction, it has a special advantage for studying the $\Delta^*$
resonance since there is no contribution from isospin $1/2$ nucleon
resonances because of the isospin and charge conservations. For the
$\Lambda p \to \Lambda p \pi^0$ reaction, it is a very good isospin
one filter for studying $\Sigma^*$ resonances decaying to $\pi
\Lambda$, and provides a useful tool for testing $\Sigma^*$ baryon
models.

Both on the experimental and theoretical sides, the nucleon and
$\Delta^*$ resonances around or above $2.0$ GeV and the hyperon
excited states have not been extensively
studied~\cite{Agashe:2014kda}. For example, the first $\Sigma(1193)$
excited state, $\Sigma^*(1385)$, was cataloged in the baryon
decuplet of the traditional quark models which still have some
problems for the excited baryon resonances. Recently, the
penta-quark picture~\cite{Helminen:2000jb,Jaffe:2003sg} provides the
natural explanation for the baryon states~\cite{Zou:2007mk}. Indeed,
the new observation of the heavy hidden charm baryonic $P^+_c$
states~\cite{Aaij:2015tga} has challenged the conventional wisdom
that baryons are composed of three quarks in the naive quark model.

Based on the penta-quark picture, a newly possible $\Sigma^*$ state,
$\Sigma^*(1380)$ ($J^P = 1/2^-$) was predicted around 1380
MeV~\cite{Zhang:2004xt}. Besides, another more general penta-quark
model~\cite{Helminen:2000jb} without introducing explicitly diquark
clusters also predicts this new $\Sigma^*$ state around 1405 MeV.
Obviously, it is helpful to check the correctness of penta-quark
models by studying the possible $\Sigma^*(1380)$ state. Because the
mass of this new $\Sigma^*$ state is close to the well established
$\Sigma^*(1385)$ resonance, it will make effects in the production
of $\Sigma^*(1385)$ resonance and then the analysis of the
$\Sigma^*(1385)$ resonance suffers from the overlapping mass
distributions and the common $\pi \Lambda$ decay mode.

In this paper, we will review the main results from those
theoretical studies about the $N^*(2120)$, $\Delta^*(1940)$, and the
possible $\Sigma^*(1380)$ resonances. In next section, we show the
results for the $N^*(2120)$ resonance in the $\gamma p \to K^+
\Lambda(1520)$ reaction. In section~\ref{sec:delta1940}, we study
the role of $\Delta(1940)$ in the $pp \to n K^+ \Sigma(1385)$
reaction, while in section~\ref{sec:sigma1380}, we investigate the
possible $\Sigma^*(1380)$ in the $\Lambda p \to \Lambda p \pi^0$
reaction. Finally, a short summary is presented in the last section.

\section{Study on $\gamma p \to K^+ \Lambda(1520)$ reaction} \label{sec:n2120}

For the $\gamma p \to K^+ \Lambda(1520)$ reaction, the differential
cross section, in the center of mass frame (c.m.), and for a
unpolarized photon beam reads~\cite{Wang:2014jxb},
\begin{eqnarray}
\frac{d\sigma}{d(\cos\theta_{\rm c.m.})}  &=& \frac{|\vec{k}_1^{\rm
\,\, c.m.}||\vec{p}_1^{\rm \,\,c.m.}|}{4\pi}\frac{M_N
M_{\Lambda^*}}{(W^2 - M_N^2)^2} \sum_{s_p, s_\Lambda^*,\lambda}
|T|^2,
\end{eqnarray}
with $W$ the invariant mass of the $\gamma p$ pair. Besides,
$\vec{k}_1^{\rm \,\, c.m.}$ and $\vec{p}_1^{\rm \,\, c.m.}$ are the
photon and $K^+$ meson c.m. three-momenta, and $\theta_{\rm c.m.}$
is the  $K^+$ polar scattering angle. The invariant scattering
amplitudes are defined as
\begin{equation}
-iT_i=\bar u_\mu(p_2,s_{\Lambda^*}) A_i^{\mu \nu} u(k_2,s_p)
\epsilon_\nu(k_1,\lambda),
\end{equation}
where $u_\mu$ and $u$ are dimensionless Rarita-Schwinger and Dirac
spinors for final $\Lambda(1520)$ and the initial proton,
respectively, while $\epsilon_\nu(k_1,\lambda)$ is the photon
polarization vector. Besides, $s_p$ and $s_{\Lambda^*}$ are the
baryon polarization variables. The sub-index $i$ stands for the
contact, $t$-channel $K^-$ exchange, $s$-channel nucleon pole and
$N^*$ resonance terms~\cite{Xie:2010yk}, and the $u$-channel
$\Lambda(1115)$ contribution~\cite{Xie:2013mua}.
\begin{figure}[htbp]
\begin{center}
\includegraphics[scale=1.]{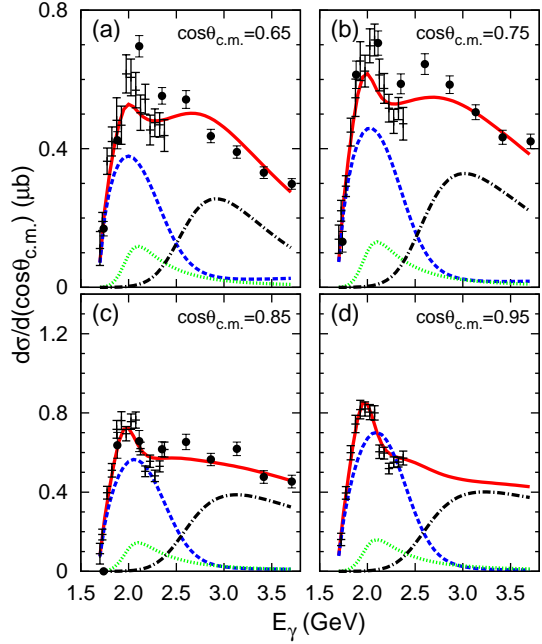}
\caption{(Color online) The $\gamma p \to K^+ \Lambda (1520)$
differential cross section as a function of the LAB frame photon
energy for different c.m. $K^+$
polar angles.} \label{dcs-leps}%
\end{center}
\end{figure}

In Fig.~\ref{dcs-leps}, we show the differential cross section as a
function of the LAB frame photon energy and for different forward
c.m. $K^+$ angles. The experimental data are taken from
LEPS~\cite{leps2} and CLAS~\cite{Moriya:2013hwg}. The blue-dashed
curve stands for the contributions from $t$-channel $K^-$ exchange,
$s$-channel nucleon pole and $u$-channel $\Lambda(1115)$ terms. The
black-dash-dotted curves stand for the contribution from the Reggeon
exchange mechanism. The green-dotted lines show the contribution of
the $N^*(2120)$ resonance term, while the red-solid lines stand for
the full contributions. One can see that our results can describe
the experimental data very well.

\section{Study on the $pp \to n K^+ \Sigma^+(1385)$ reaction}
\label{sec:delta1940}

The full invariant scattering amplitude for the $p p \to n K^+
\Sigma^+(1385)$ reaction is composed of two parts corresponding to
the $s$-channel $\Delta^*(1940)$ resonance, and $u$-channel
$\Lambda(1115)$ hyperon pole, which are produced by the
$\pi^+$-meson exchanges, ${\cal M} = {\cal M}_s + {\cal M}_u$ (more
details can be found in Ref.~\cite{Xie:2014kja}). Here we give
explicitly the amplitude ${\cal M}_s$, as an example,
\begin{eqnarray}
{\cal M}_s & = & \frac{\sqrt{2} g_{\pi NN} g_{\pi N
\Delta^*}}{m_{\pi}} F^{N N}_{\pi}(k^2_{\pi}) F^{\Delta^*
N}_{\pi}(k^2_{\pi}) F_s(q_s^2) G_{\pi}(k^2_{\pi}) \bar{u}^{\mu}
(p_4,s_4)
(-\frac{g_1}{m_K} \Slash p_5 g_{\mu \rho} + \frac{g_2}{m^2_K}p_{5 \mu} p_{5\rho}) \times  \nonumber\\
&&  G^{\rho \sigma}_{\Delta^*}(q_s) k_{\pi \sigma} \gamma_5
u(p_1,s_1)\bar{u}(p_3,s_3) \gamma_5 u(p_2,s_2)   + (\text {exchange
term with } p_1 \leftrightarrow p_2), \label{ppampms}
\end{eqnarray}
where $s_i~(i=1,2,3)$ and $p_i~(i=1,2,3)$ represent the spin
projection and 4-momenta of the two initial protons and final
neutron, respectively. While $p_4$ and $p_5$ are the 4-momenta of
the final $\Sigma^+(1385)$ and $K^+$ meson, respectively. And $s_4$
stands for the spin projection of $\Sigma^+(1385)$. In
Eq.~(\ref{ppampms}), $k_{\pi} = p_2 - p_3$ and $q_s = p_4 + p_5$
stand for the 4-momenta of the exchanged $\pi^+$ meson and
intermediate $\Delta^{*}(1940)$ resonance. The $G_{\pi}(k^2_{\pi})$
and $G^{\rho \sigma}_{\Delta^*}(q_s)$ are the pion and
$\Delta^*(1940)$ propagators, which have the forms as,
\begin{eqnarray}
G_{\pi}(k^2_{\pi}) &=& \frac{i}{k_{\pi}^2-m^2_{\pi}}, \\
G^{\rho \sigma}_{\Delta^*}(q_s) &=&  \frac{i(\Slash q_s +
M_{\Delta^*})}{q^2_s - M^2_{\Delta^*} +
iM_{\Delta^*}\Gamma_{\Delta^*}} \left ( -g^{\rho \sigma} +
\frac{1}{3}\gamma^{\rho}\gamma^{\sigma} +
\frac{2}{3M^2_{\Delta^*}}q_s^{\rho}q_s^{\sigma}  +
\frac{1}{3M_{\Delta^*}}(\gamma^{\rho}q_s^{\sigma} -
\gamma^{\sigma}q_s^{\rho}) \right ),
\end{eqnarray}
where $M_{\Delta^*}$ and $\Gamma_{\Delta^*}$ are the mass and total
decay width of the $\Delta^*(1940)$ resonance, respectively.

Furthermore, we need also the relevant off-shell form factors for
$\pi NN$ and $\pi N \Delta^*$ vertexes, which have been already
included in the amplitude of Eq.~(\ref{ppampms}), and we take them
as,
\begin{eqnarray}
F^{NN}_{\pi}(k^2_{\pi}) = \frac{\Lambda^2_{\pi}-m_{\pi}^2}
{\Lambda^2_{\pi}- k_{\pi}^2}, ~~~~ F^{\Delta^* N}_{\pi}(k^2_{\pi}) =
\frac{\Lambda^{*2}_{\pi} -
m_{\pi}^2}{\Lambda^{*2}_{\pi}-k_{\pi}^2},~~~~ F_s(q_s^2) =
\frac{\Lambda^4_s}{\Lambda^4_s + (q_s^2-M_{\Delta^*}^2)^2},
\end{eqnarray}
with $k_{\pi}$ the 4-momentum of the exchanged $\pi$ meson. The
cutoff parameters are taken as $\Lambda_{\pi} = \Lambda^*_{\pi}$ =
1.1 GeV and $\Lambda_s = 0.9$ GeV, with which the experimental data
on $p p \to n K^+ \Sigma^+(1385)$ reaction can be reproduced.

Then, we can easily obtain the total cross sections for $pp \to n
K^+ \Sigma^+(1385)$ as,
\begin{eqnarray}
d\sigma  &=& \frac{m^2_p m_n m_{\Sigma^+(1385)}}{256 \pi^5
\sqrt{(p_1\cdot p_2)^2-m^4_p}} \sum_{s_1, s_2,s_3, s_4} |{\cal M}|^2
\frac{d^{3} p_{3}}{E_{3}} \frac{ d^{3} p_4}{E_4} \frac{d^{3}
p_5}{E_5} \delta^4 (p_{1}+p_{2}-p_{3}-p_{4}-p_5). \label{ppdcs}
\end{eqnarray}

\begin{figure}[htbp]
\begin{center}
\includegraphics[scale=0.45]{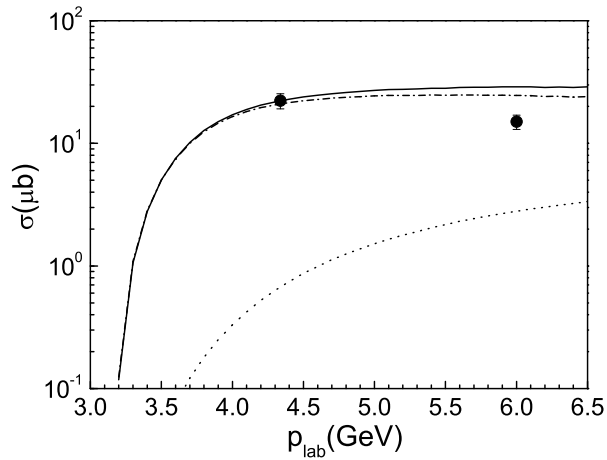}
\caption{Total cross sections vs beam energy ${\rm p_{lab}}$ of
proton for the $pp \to n K^+\Sigma^+(1385)$ reaction.} \label{pptcs}
\end{center}
\end{figure}

The total cross section versus the beam momentum are shown in
Fig.~\ref{pptcs}. The dotted, and dash-dotted lines stand for
contributions from $\Lambda(1115)$ and $\Delta^{*}(1940)$ resonance,
respectively. Their total contributions are shown by the solid line.
From Fig.~\ref{pptcs}, we can see that the contribution from the
$\Delta^*(1940)$ resonance is predominant in the whole considered
energy region. For comparison, we also show the experimental
data~\cite{Agakishiev:2011qw,Klein:1970ri} in Fig.~~\ref{pptcs},
from where we can see that our predictions for the total cross
sections of $pp \to n K^+ \Sigma^+(1385)$ reaction are in agreement
with the experimental data.

\section{Study on the $\Lambda p \to \Lambda p \pi^0$ reaction}
\label{sec:sigma1380}

\begin{figure}[htbp]
\begin{center}
\includegraphics[scale=0.45]{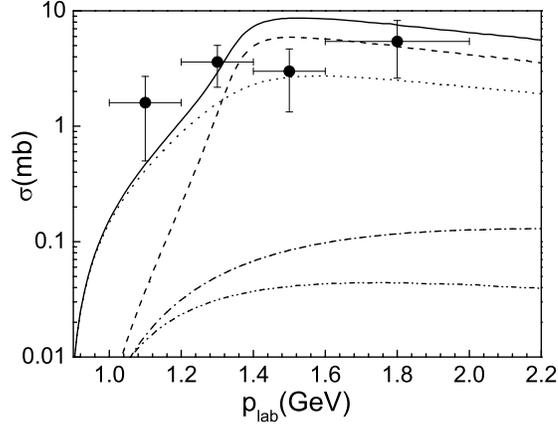}
\caption{The total cross sections vs the beam momentum p$_{\rm lab}$
for the $\Lambda p \to \Lambda p \pi^0$ reaction.}
\label{Fig:tcsres}
\end{center}
\end{figure}

The theoretical results (see more details in
Ref.~\cite{Xie:2014zga}) of the total cross section for the $\Lambda
p \to \Lambda p \pi^0$ reaction versus the beam momentum are shown
in Fig.~\ref{Fig:tcsres}, where the contributions of
$\Sigma^*(1385)$ resonance, $\Sigma^*(1380)$ state, nucleon pole and
$\Sigma(1193)$ pole to the energy dependence of the total cross
section are shown by dashed, dotted, dash-dotted, and
dash-dot-dotted curves, respectively. Their total contribution is
depicted by the solid line. It is clear that the contributions from
the $\Sigma^*(1380)$ state and $\Sigma^*(1385)$ resonance dominate
the total cross section at beam momenta below and above $1.3$ GeV,
respectively, while the contributions of nucleon and $\Sigma(1193)$
pole are small and can be neglected.

The $\Sigma^*(1385)$ resonance with spin-parity $3/2^+$ decays to
$\pi \Lambda$ in relative $P$-wave and is suppressed at low
energies. It can not reproduce the near threshold enhancement for
the $\Lambda p \to \Lambda p \pi^0$ reaction. On the contrary, the
possible $\Sigma^*(1380)$ state with $J^P=1/2^-$ is decaying to $\pi
\Lambda$ in relative $S$-wave, which will give enhancement at the
near threshold. As one can see in Fig.~\ref{Fig:tcsres}, thanks to
the contribution from $\Sigma^*(1380)$ state, we can reproduce the
experimental data taken from Ref.~\cite{Kadyk:1971tc} for all of the
beam energies. Thus, we find a natural source for the near threshold
enhancement of the $\Lambda p \to \Lambda p \pi^0$ reaction coming
from the possible $\Sigma^*(1380)$ state which decays to $\pi
\Lambda$ in the $S$-wave.

\section{Summary} \label{sec:summary}

The combination of effective Lagrangian approach and isobar model is
an important theoretical tool in describing the various processes in
the region of baryon resonance production. In this paper, we have
shown the results on the studies of the $N^*(2120)$,
$\Delta^*(1940)$, and the possible $\Sigma^*(1380)$ resonances in
the $\gamma p \to K^+ \Lambda(1520)$, $pp \to n K^+ \Sigma(1385)$,
and $\Lambda p \to \Lambda p \pi^0$ reactions. It is shown that when
the contributions from these baryonic states were considered, the
current experimental measurement could be well reproduced. In
addition, we also demonstrate that the angular distributions provide
direct information of these reaction, hence could be useful for the
investigation of those states and may be tested by future
experiments.

\section*{Acknowledgments}

We would like to thank J. Nieves and B.S. Zou for collaborations on
relevant issues reported here. This work is partly supported by the
National Natural Science Foundation of China under Grant Nos.
11475227 and 11505158. It is also supported by the Open Project
Program of State Key Laboratory of Theoretical Physics, Institute of
Theoretical Physics, Chinese Academy of Sciences, China
(No.Y5KF151CJ1).

\end{document}